\documentclass[12pt]{article}
\usepackage{epsfig,amsfonts,amssymb}
\usepackage{hyperref}
\usepackage{cite}
\input epsf.sty
\usepackage{cite}

\def\ZZZ{{\hbox{ Z\kern-1.6mm Z}}}
\def\RRR{{\hbox{ R\kern-2.4mm R}}}
\def\CCC{{\hbox{ C\kern-2.0mm C}}}
\def\zzz{{\hbox{z\kern-1mm z}}}

\newcommand{\vt}{\vartheta}

\newcommand{\qeq}{{\hbox{=\kern-2.3mm ? \kern.5mm }}}
\renewcommand{\qeq}{=}

\newcommand{\II}{{\cal I}}

\newcommand{\GG}{{\cal G}}

\newcommand{\FF}{{\cal F}}
\newcommand{\JJ}{{\cal J}}

\newcommand{\MM}{{\cal M}}
\newcommand{\CC}{{\cal C}}

\newcommand{\wt}{\widetilde}

\newcommand{\NN}{{\cal N}}

\newcommand{\tI}{\wt\II}

\newcommand{\be}{\begin{equation}}
\newcommand{\ee}{\end{equation}}
\newcommand{\ben}{\begin{eqnarray}\displaystyle}
\newcommand{\een}{\end{eqnarray}}

\newcommand{\bea}[1]{\begin{eqnarray}\label{#1} }
\newcommand{\eea}{\end{eqnarray}}

\newcommand{\refb}[1]{(\ref{#1})}

\newcommand{\sectiono}[1]{\section{#1}\setcounter{equation}{0}}

\def\one{{\hbox{ 1\kern-.8mm l}}}
\def\zero{{\hbox{ 0\kern-1.5mm 0}}}

\begin{document}

\baselineskip 24pt

\begin{center}
{\Large  \bf
Discrete Information from CHL Black Holes}

\end{center}

\vskip .6cm
\medskip

\vspace*{4.0ex}

\baselineskip=18pt

\centerline{\large \rm   Ashoke Sen }

\vspace*{4.0ex}

\centerline{\large \it Harish-Chandra Research Institute}
\centerline{\large \it  Chhatnag Road, Jhusi,
Allahabad 211019, India}
\centerline{and}
\centerline{\large \it 
LPTHE, Universite Pierre et Marie Curie, Paris 6}
\centerline{\large \it 
4 Place Jussieu,  75252 Paris Cedex 05, France}

\vspace*{1.0ex}
\centerline{E-mail:  sen@mri.ernet.in, ashokesen1999@gmail.com}

\vspace*{5.0ex}

\centerline{\bf Abstract} \bigskip

$AdS_2/CFT_1$ correspondence predicts that the logarithm of a
$\ZZZ_N$ twisted index over states carrying a fixed set of
charges grows as $1/N$ times the entropy of the
black hole carrying the same set of charges.
In this paper we verify this explicitly by calculating the
microscopic $\ZZZ_N$ twisted index for a class of states in
the CHL models. This demonstrates that black holes carry more
information about the microstates than just the total degeneracy.

\vfill \eject

\baselineskip=18pt

\tableofcontents

\sectiono{Introduction and Summary} \label{sint}

CHL models\cite{9505054,9506048} 
in four dimensions with $\NN=4$ 
supersymmetry have proved
to be
a rich arena for studying the physics of black 
holes\cite{0510147,0602254,0603066,0605210,0607155,0609109,
0612011}. On
the one hand they have as much supersymmetry and hence as much
control as the toroidally compactified heterotic string theory. On
the other hand they have different effective actions beyond the
supergravity approximation and hence make different predictions for the
entropy of  BPS black holes beyond the leading order result of
\cite{9507090,9512031}. Thus
they provide us with more data points at which we can compare the
macroscopic and microscopic predictions for the black hole entropy.
This comparison has been remarkably successful at the level of four
derivative corrections to the effective action, reproducing
complicated non-trivial functional dependence of the entropy on
the charges on both sides.\footnote{We should add 
a note of caution that
this comparison requires us to make assumption of certain 
non-renormalization results which have not been proven. In particular
it assumes that at the level of four derivative terms the Gauss-Bonnet
terms (or their supersymmetric completion
given in \cite{9602060,9603191,9812082,0007195})
in the action are sufficient to calculate the correction
to the black hole entropy. The analysis in this paper does not
require us to make any such assumption.} Indeed, most of the
results on black holes in heterotic string theory on 
$T^6$\cite{9607026,0412287,0505094,0506249,0508174,0605210,0705.1433,
0802.0544,0802.1556,0803.2692}
have now been generalized to the case of CHL models.

In this paper we shall make use of the CHL model to explore another
aspect of black holes. Based on $AdS_2/CFT_1$
correspondence\cite{0809.3304,0903.1477} it was argued in
\cite{0911.1563} that if a theory has a $\ZZZ_N$ 
symmetry that cannot be
regarded as part of a $U(1)$ gauge transformation, 
and if we pick a black hole carrying $U(1)$ charges which are
invariant under this $\ZZZ_N$ transformation, then the
logarithm of the trace of the $\ZZZ_N$ generator over the
microstates of the black hole grows as  $1/N$ times the 
entropy of the
black hole.\footnote{For a $\ZZZ_N$ group that can
be regarded as a subgroup of a spontaneously broken $U(1)$
gauge group, the possibility of hair modes containing information
about the $\ZZZ_N$ quantum numbers was explored in
\cite{wil1,wil2}. In contrast the $\ZZZ_N$ groups 
we discuss here cannot be
regarded as a subgroup of a spontaneously broken $U(1)$
symmetry. Also our goal here is quite different from the one
of \cite{wil1,wil2}.}
This can be made more concrete for BPS
black holes in
supersymmetric string theories by working with protected
helicity trace index.
In the context of $\NN=4$ supersymmetric string theories in four
dimensions the relevant twisted index is the 6th helicity 
trace index\cite{9708062,9708130,0911.1563}:
\be \label{ehe1}
B^g_6(\vec q) = {1\over 6!}\,
Tr_{\vec q} \left\{(-1)^{2h} (2h)^6\, g\right\}\, ,
\ee
where  the trace is taken over all states carrying a fixed set
of charges $\vec q$, $h$ is the third component of the angular
momentum of the state in its rest frame, and $g$ is the generator
of a $\ZZZ_N$ symmetry which leaves $\vec q$ invariant. 
This index receives contribution from
$1/4$ BPS states in this theory which describe large black holes
with near horizon $AdS_2\times S^2$ geometry. In this case the
analysis 
of \cite{0911.1563} applies and tells us that
\be \label{emacpred}
\left|B^g_6(\vec q)\right|\sim \exp[S_{BH}(\vec q)/N]\, ,
\ee 
where $S_{BH}(\vec q)$ is the entropy of an extremal black hole
carrying charge $\vec q$. 
We shall not review the arguments of \cite{0911.1563} here; but the
central idea is that in computing the contribution to \refb{emacpred}
from the horizon of the black hole the leading saddle point corresponding
to the $AdS_2\times S^2$ near horizon geometry does not contribute.
However a $\ZZZ_N$ orbifold of 
$AdS_2\times S^2$\cite{0810.3472,0903.1477,0904.4253},
whose asymptotic geometry coincides with that of the original
near horizon geometry of the black hole, 
contributes and gives the answer
$\exp[S_{BH}/N]$ in the semiclassical limit.

While \refb{emacpred} follows almost trivially from the
$AdS_2/CFT_1$ correspondence, it
is quite striking from the point of view of the
microscopic theory. 
For large black holes the right hand side of \refb{emacpred} is much
smaller than the untwisted helicity trace index carrying the same
charges, since the latter is given by $\exp[S_{BH}(\vec q)]$.
What this tells us is that in a given charge sector
the microstates
of different $g$ eigenvalues come in almost equal numbers so
that the sum weighted by $g$ is much smaller than the total number
of states.
This was explicitly verified in \cite{0911.1563} by
deriving the microscopic formula for this twisted index in
toroidally compactified heterotic and type II string 
theories and then studying their 
asymptotic behaviour.\footnote{Even though type II 
string theory on $T^6$ has
$\NN=8$ supersymmetry, only an $\NN=4$ subgroup of this
commutes with $g$. Thus effectively we can analyze it in the same
way as in an $\NN=4$ supersymmetric theory.}

Given the unusual nature of this macroscopic prediction, it is
important to test this in as many examples as possible. 
In
this paper we shall verify this in the context of CHL models.
The construction of the CHL models that we shall analyze
proceeds as follows. We begin with
type IIB string theory on $\MM\times S^1\times \wt S^1$ where
$\MM$ is either K3 or $T^4$ and go to a special subspace of
the moduli space of $\MM$ where the theory has a geometric
$\ZZZ_M\times \ZZZ_N$ symmetry
that commutes with 16 supersymmetry generators of the theory.
An extensive list of possible symmetries of this type can be
found in \cite{9508144,9508154}. Note also that a $Z_{MN}$
group with $M$ and $N$ relatively prime can be considered as
a $Z_M\times Z_N$ group for the purpose of our analysis. 
Let us denote by $g_M$ and $g_N$ the generators of $\ZZZ_M$ and
$\ZZZ_N$ respectively. We now take an orbifold of this theory by
a symmetry that involves $1/M$ unit of translation along the circle
$S^1$ accompanied by the transformation $g_M$. 
This gives a theory
with $\NN=4$ supersymmetry in four dimensions and the $\ZZZ_N$ group
generated by $g_N$ is a symmetry of this theory. 
We now consider a $g_N$ invariant
charge vector $\vec q$ 
in this theory and define
the index
\be \label{ek1}
d(\vec q) = -{1\over 6!}\,
Tr_{\vec q} \left(e^{2\pi i h} (2h)^6 g_N\right)\, ,
\ee
where the trace is taken over all states carrying the charge 
$\vec q$.
Eq.\refb{emacpred}
now translates to
\be \label{ek2}
\left|d(\vec q)\right|\sim \exp[S_{BH}(\vec q)/N]\, ,
\ee
for large charges.
Our goal will be to verify this by explicit computation
of $d(\vec q)$ in the microscopic theory.

Since the explicit counting of states involves
technical details, we shall take this opportunity to summarize
the results of our analysis.
We use a convention in
which the coordinate radius of the original circle $S^1$
before orbifolding is $2\pi M$ so that the orbifold action involves
translation by $2\pi$ along $S^1$ accompanied by $g_M$. In this
convention the
minimum amount of momentum along $S^1$ is $1/M$. 
We focus on states carrying one unit of KK monopole charge associated
with the circle $\wt S^1$, one unit of D5-brane charge wrapped on
$\MM\times S^1$, $Q_1$ units of D1-brane charge wrapped on
$S^1$, left-moving momentum $n/M$ along $S^1$ and
$J$ units of momentum along $\wt S^1$, and define
\be \label{ek3}
Q^2 = 2n/M, \qquad P^2 = 2Q_1, \qquad Q.P=J\, .
\ee
In this case our result of $d(\vec q)$ is given by
\be \label{ek4}
d(\vec q) = {1\over M} (-1)^{Q.P+1} \int_{\CC} \, d\rho d\sigma
d v e^{-\pi i(M\rho Q^2 + \sigma P^2/M + 2 v Q.P)} 
{1\over \wt\Phi(\rho,\sigma,v)}\, .
\ee
Here $\CC$ is a three real dimensional subspace of the three
complex dimensional space labelled by $(\rho=\rho_1+i
\rho_2, \sigma=\sigma_1+i\sigma_2, v=v_1+iv_2)$ given by
\ben \label{ek5}
&& \rho_2=M_1, \qquad \sigma_2 = M_2, \qquad v_2 = M_3,
\nonumber \\
&& 0\le\rho_1 \le 1, \qquad 0\le\sigma_1\le M, \qquad
0\le v_1 \le 1\, ,
\een
$M_1$, $M_2$, $M_3$ being large but fixed positive numbers
satisfying
\be \label{ek6}
M_1 M_2 > M_3^2\, .
\ee
The function $\wt\Phi(\rho,\sigma,v)$ is a modular form
of a subgroup of $Sp(2,\ZZZ)$, given by
\ben \label{ek7}
\wt\Phi(\rho,\sigma,v) &=&  e^{2\pi i (\wt\alpha \rho
+\wt\gamma\sigma + \wt\beta v)} \,
\prod_{b=0}^1 \prod_{r=0}^{N-1}\prod_{r'=0}^{M-1}
\prod_{k\in\zzz+{r'\over M}, l\in\zzz, j\in 2\zzz+b\atop
k,l\ge 0, j<0 \, for\, k=l=0}
\left[ 1 - e^{2\pi i r/N} \, e^{2\pi i (k\sigma +l\rho + j v)}
\right]^a\nonumber \\
a &\equiv& \sum_{s=0}^{N-1} \sum_{s'=0}^{M-1} e^{-2 
\pi i (s'l / M +
rs/N)} c_b^{(0,s;r',s')}(4kl - j^2)\, ,
\een
where the coefficients $c_b^{(r,s;r',s')}$ are defined via the 
equation:
\ben \label{ek8}
&&\sum_{b=0}^1 \sum_{j\in 2\zzz+b, n\in \zzz/MN} c_b^{(r,s;r',s')}
(4n - j^2) e^{2\pi i (n\tau + j z)} \nonumber \\
&=& {1\over MN} Tr_{RR;g_M^{r'} g_N^r} \left (g_M^{s'} g_N^{s} 
(-1)^{J_L+J_R} e^{2\pi i (\tau L_0-\bar\tau \bar L_0)}
e^{2\pi i J_L z} \right)\, .
\een
The trace is taken over all the $g_M^{r'} g_N^{r}$ twisted
RR sector states in the (4,4) superconformal  CFT$_2$
with target space $\MM$.
$L_0$ and $\bar L_0$ are the left and right-moving Virasoro
generators and $J_L/2$ and $J_R/2$ are the generators
of the $U(1)_L\times U(1)_R$ subgroup of the $SU(2)_L\times 
SU(2)_R$
R-symmetry group of this CFT$_2$. 
An algorithm for explicitly 
computing the right hand side of \refb{ek8}
has been outlined in appendix \ref{sb}.
The coefficients 
$\wt\alpha$, $\wt\beta$, $\wt\gamma$ are given by
\ben \label{ek9}
\wt\alpha &=& {1\over 24M} Q_{0,0} -{1\over 2M}
\sum_{s'=1}^{M-1} Q_{0,s'} {e^{-2\pi i s'/M}
\over (1- e^{-2\pi i s'/M}
)^2}\, , \nonumber \\
\wt\beta &=& 1\nonumber \\
\wt\gamma &=& {1\over 24 M}\, \chi(\MM) = {1\over 24 M}\, Q_{0,0}\, ,
\een
where
\be \label{edefqrs}
Q_{r',s'} = M N \left( c_0^{(0,0;r',s')}(0)
+ 2  c_1^{(0,0;r',s')}(-1) \right)\, .
\ee
Eqs. \refb{ek8}, \refb{ek9} define all the quantities which appear
in the definition of $\wt\Phi$. The only ambiguity that
remains in computing the right hand side of \refb{ek4} is the choice
of the integration contour encoded in the choice of $(M_1, M_2, M_3)$.
As is well known by now, this ambiguity is related to the phenomenon
of wall crossing\cite{0702141,0702150,0705.3874,0706.2363}. 
Different choices of $M_1$, $M_2$ and $M_3$ give
the value of $d(\vec q)$ for different values of the asymptotic moduli.
However the ambiguity in the value of $d(\vec q)$ that it introduces is
sufficiently small so as not to affect our analysis, and hence we
shall ignore it in our subsequent discussion.

Given the result \refb{ek4} for $d(\vec q)$ we can find its behaviour for
large $Q^2$, $P^2$ and $Q.P$ by standard 
method\cite{9607026,0412287,0510147,0605210}. 
The result is that
$d(\vec q)$ behaves as
\be \label{ek10}
d(\vec q)\sim \exp\left[ \pi\sqrt{Q^2 P^2 - (Q.P)^2}/N\right]\, .
\ee
Since in this limit a black hole of charge $\vec q$ has 
entropy\cite{9507090,9512031}
\be \label{ek11}
S_{BH}(\vec q) \simeq \pi\sqrt{Q^2 P^2 - (Q.P)^2}\, ,
\ee
we see that the microscopic result \refb{ek10} is in perfect
agreement with the macroscopic prediction \refb{ek2}.

Finally we would like to remark that even though we have presented our
analysis for the index $Tr((-1)^{2h} (2h)^6 g_N)$, we can repeat the
analysis with $g_N$ replaced by $(g_N)^b$ for any integer $b$. 
In this case
the role of $N$ is played by the order of $(g_N)^b$, and in 
all the formul\ae\
we simply have to replace $g_N$ by $(g_N)^b$. This in turn allows
us to compute the index $Tr((-1)^{2h}(2h)^6)$ 
for states carrying a definite $g_N$
eigenvalue $e^{2\pi i a/N}$ using the combination
\be \label{ecomb}
{1\over N} \, \sum_{b=0}^{N-1} e^{-2\pi i ab/N} \, 
Tr((-1)^{2h} (2h)^6 (g_N)^b)\, .
\ee
Thus our result can also be interpreted as the agreement between the
macroscopic and the microscopic results for the helicity trace
index over states carrying a definite $g_N$ charge.

\sectiono{The counting} \label{scount}

The counting of states of the D1-D5-KK monopole system proceeds
as in \cite{0605210,0607155,0609109,0708.1270}. 
We take the circle $S^1$ to be large compared to the size of $\MM$
and regard the world-volume theory as a 1+1
dimensional field theory living on $S^1$.
Denoting by $d(Q_1,n,J)$ the twisted index \refb{ek1} of
states carrying charge labeled by $(Q_1,n,J)$ in the convention of
\S\ref{sint}, we define
\be \label{edefz}
Z(\rho,\sigma, v) = \sum_{Q_1,n,J} e^{2\pi i (Q_1\sigma/M + n\rho
+ vJ)} (-1)^J d(Q_1,n,J)\, .
\ee
Proving \refb{ek4} is now equivalent to proving that $Z=-1/\wt\Phi$. 

The twisted partition function $Z$ is given by the product of
three twisted
partition functions, -- that of the excitations living
on the KK monopole, that of the dynamics describing the overall
motion of the D1-D5 system in the background of the KK monopole
and that of the motion of the D1-brane along the D5-brane.
For each system we must keep the right-movers in their ground state
and excite the left-movers in order to preserve 
supersymmetry.\footnote{Here and elsewhere left-moving modes
will refer to modes carrying momentum along the negative
$S^1$ direction. Thus left-moving momentum $n$ will indicate
momentum $-n$ along $S^1$.} 
Since the fermion zero modes associated
with the broken supersymmetries are automatically removed 
while computing the helicity trace (which is $B_6$ in this case
since the system breaks 12 of the 16 supersymmetries), we shall ignore
their contribution during the rest of our analysis.

We begin by analyzing the
partition function of the KK monopole. 
The massless bosonic modes on the world-volume of the
KK monopole arise from the motion along the three
transverse directions and the components of the $p$-form fields
along the product of the harmonic $(p-2)$-forms of $\MM$ and the
harmonic 2-form of the Taub-NUT space. The massless fermions
are the goldstinos associated with the supersymmetries broken by
the Kaluza-Klein monopole.
In general one can show that the left-moving bosons and fermions
are in one to one correspondence with the even and odd degree
harmonic forms on $\MM$\cite{0708.1270}.  Furthermore their
$(g_M,g_N)$ quantum numbers are also given by the
$(g_M,g_N)$ eigenvalues of the harmonic forms on $\MM$. 
Since the harmonic $(p,q)$
forms on $\MM$ are in one to one
correspondence with the RR sector ground states in the
supersymmetric $\sigma$-model with target space 
$\MM$ carrying $L_0=\bar L_0=0$, 
$J_L=(p-1)$, $J_R=(q-1)$, it follows
from \refb{ek8} that 
the number of left-moving
bosons minus the number of left-moving
fermions on the KK monopole world-volume,
carrying $g_N$ quantum number $e^{2\pi i r/N}$ and
$g_M$ quantum number $e^{2\pi i k'/M}$, is given
by\cite{0609109,0708.1270}
\ben \label{ea1}
&& {1\over MN}\,
\sum_{s=0}^{N-1}\sum_{s'=0}^{M-1} e^{-2\pi i r s/N} 
e^{-2\pi i k's'/M} Tr_{RR;I} \left[ (-1)^{J_L+J_R} \, g_M^{s'}
g_N^s\, 
\delta_{L_0,0}\, \delta_{\bar L_0,0}\right]\nonumber \\
&=&
\sum_{s=0}^{N-1}\sum_{s'=0}^{M-1} e^{-2\pi i r s/N} 
e^{-2\pi i k's'/M} \left(c_0^{(0,s;0,s')}(0) + 2 c_1^{(0,s;
0,s')}(-1)
\right)\, .
\een
In arriving at \refb{ea1} we have used the fact that
$c_b^{(r,s;r',s')}(u)=0$ for $u<-1$.
Now consider a mode carrying $g_M$ eigenvalue
$e^{2\pi i k'/M}$ and left-moving momentum $l/M$
along $S^1$. The requirement of invariance under the simultaneous
action of $g_M$ and $2\pi$  translation along
$S^1$ gives us the requirement that $l=k'$ mod $M$. 
Furthermore
the contribution to the twisted index from these states will be
weighted by the $g_N$ eigenvalue
$e^{2\pi i r/N}$. Thus the net contribution to the
partition function from these modes is given by
\be \label{ea2}
Z_{KK} = e^{-2\pi i\wt\alpha\rho}\,
\prod_{r=0}^{N-1} \prod_{l=1}^\infty \left(1 - e^{2\pi i r/N}
e^{2\pi i l\rho}\right)^{-\sum_{s=0}^{N-1}\sum_{s'=0}^{M-1} 
e^{-2\pi i r s/N} 
e^{-2\pi i ls'/M} \left(c_0^{(0,s;0,s')}(0) + 2 c_1^{(0,s;
0,s')}(-1)
\right)}\, .
\ee
Here the term $e^{-2\pi i\wt\alpha\rho}$ reflects the effect of the
momentum carried by the ground state of the Kaluza-Klein monopole.
The analysis of \cite{0609109,0708.1270} gives
\ben \label{ea3}
\wt\alpha &=& {1\over 24M} Q_{0,0} -{1\over 2M}
\sum_{s'=1}^{M-1} Q_{0,s'} {e^{-2\pi i s'/M}\over (1- e^{-2\pi i s'/M}
)^2}\, ,
\nonumber \\
&& Q_{r',s'} \equiv M N \left( c_0^{(0,0;r',s')}(0)
+ 2  c_1^{(0,0;r',s')}(-1) \right)\, .
\een

Next we turn to the dynamics of the overall motion of the D1-D5
system in the KK monopole background. 
The dynamics in the transverse direction
is independent of whether we are working with $K3$ or $T^4$.
Furthermore these modes do not carry any $g_N$ or $g_M$
quantum numbers; thus the contribution from these modes to the
partition function is universal. The result 
is\cite{0609109,0708.1270}
\ben \label{ere1}
&& -  e^{-2\pi i v} \left( 1 - e^{-2\pi i v}\right)^{-2}\nonumber \\
&&\prod_{l\in M\zzz\atop l>0} \{ (1-e^{2\pi i l\rho})^4 
(1-e^{2\pi i l\rho+
2\pi i v})^{-2}
(1-e^{2\pi i l\rho-
2\pi i v})^{-2}\}\, .
\een
The first line represents the contribution from the zero mode
dynamics that binds the D1-D5 system to the KK 
monopole\cite{pope,9912082,0605210}, 
and the second line represents the contribution from the
oscillators. 
The last two terms in the second line of \refb{ere1} represent the
contribution from the four left-moving bosonic modes representing
transverse oscillation of the D1-D5 system whereas the first factor
in the same line represents contribution from the left-moving
fermionic modes.\footnote{These left-moving bosonic and fermionic
modes, as well as those which contribute to \refb{etw1}, are
paired by the unbroken superysymmetry transformations on the
D1-D5 world volume in flat space-time
which commute with $\ZZZ_M\times \ZZZ_N$,
are charged under the $SU(2)_L$ subgroup of the transverse rotation
group,
and act on the left-movers. Eventually when we place this system
in the background of KK monopole this supersymmetry is broken since
in the full system there is no supersymmetry acting on the left-movers.
However this is still useful for determining the quantum numbers of
the fermions from the known quantum numbers of the bosonic 
modes\cite{0609109,0708.1270}. \label{f1}}
In arriving at \refb{ere1} one needs to use the fact that
in the presence of the KK monopole background, the momentum along
$\wt S^1$ appears as the angular momentum 
$2J_L$  for the D1-D5 system where $J_L$ is the
generator of the $U(1)_L\subset SU(2)_L$ subgroup
of the rotation
group in transverse space\cite{0503217}. The $v$ dependence of
\refb{ere1} then follows from the fact that the bosonic modes,
transforming as a vector of the transverse rotation group
$SU(2)_L\times SU(2)_R$, carry
$J_L=\pm 1$ while the fermionic modes are neutral under 
$U(1)_L$ as a consequence of footnote \ref{f1}.

For $\MM=T^4$ we also have four 
additional bosonic modes arising from the
Wilson lines on the D5-brane along $T^4$ and four additional
fermionic modes. In order to find the
contribution to the partition function from these modes we need to
know the action of $g_M$ and $g_N$ on these
modes. If $z_1$ and $z_2$ denote the complex coordinates on
$T^4$ then in order to preserve supersymmetry both $g_M$ and
$g_N$ must act as equal and opposite rotation of $z_1$ and $z_2$,
possibly
accompanied
by shifts. We shall assume
for definiteness that $g_M$ and $g_N$ induce respectively
$2\pi/M$ and $2\pi/N$ rotations on these coordinates:
\ben \label{e2pi}
g_M &:&  (dz_1, dz_2)\to \left( e^{2\pi i/M}dz_1,
e^{-2\pi i/M}dz_2\right) \, , \nonumber \\
g_N &:&  (dz_1, dz_2)\to \left( e^{2\pi i/N}dz_1,
e^{-2\pi i/N}dz_2\right)\, .
\een
\refb{e2pi}  represents the action of $g_M$ and $g_N$ on the
Wilson line variables. Furthermore the Wilson lines are
neutral under the rotation group in the transverse space
and hence carry $J_L=0$.
The result of footnote \ref{f1} now tells us that
the
additional fermionic modes on the D1D5 system, which arise for
$\MM=T^4$, transform in the same way under $g_M$ and $g_N$,
and carry $J_L=\pm 1$ uncorrelated
with their $(g_M,g_N)$ quantum numbers\cite{0609109,0708.1270}.
Thus
the contribution from these additional modes to the
twisted partition function is given by
\ben \label{etw1}
&& \prod_{l\in M\zzz+1\atop l>0} \left(1 - e^{2\pi i/N} e^{2\pi i l\rho}
\right)^{-2}\prod_{l\in M\zzz-1\atop l>0} 
\left(1 - e^{-2\pi i/N} e^{2\pi i l\rho}
\right)^{-2}\prod_{l\in M\zzz+1\atop l>0} 
\left(1 - e^{2\pi i/N} e^{2\pi i l\rho + 2\pi i v}
\right)\nonumber \\
&&\prod_{l\in M\zzz+1\atop l>0} 
\left(1 - e^{2\pi i/N} e^{2\pi i l\rho - 2\pi i v}
\right)\prod_{l\in M\zzz-1\atop l>0} 
\left(1 - e^{-2\pi i/N} e^{2\pi i l\rho+2\pi i v}
\right)
\prod_{l\in M\zzz-1\atop l>0} 
\left(1 - e^{-2\pi i/N} e^{2\pi i l\rho-2\pi i v}
\right)\, . \nonumber \\
\een
The first two factors come from the bosonic modes and the last
four factors arise from the fermionic modes whose contribution have
not already been included in \refb{ere1}.
The only new ingredient in this formula compared to that in
\cite{0609109,0708.1270} is the insertion of the factors
of $e^{\pm 2\pi i/N}$, -- these arise from the insertion of $g_N$
into the trace.

The product of \refb{ere1} and \refb{etw1} can be written in a
compact form using the coefficients $c_1^{(0,s;0,s')}(-1)$. It follows
from its definition, and the identification of the RR sector ground
states in the SCFT
with target space $\MM$ carrying $(J_L,J_R)=(p-1,q-1)$
with the harmonic $(p,q)$ forms on
$\MM$, that $MNc_1^{(0,s;0,s')}(-1)$ 
represents trace over the $(0,q)$
forms on $\MM$ weighted by $(-1)^{q} g_N^s 
g_M^{s'}$\cite{0609109,0708.1270}.
On $K3$ the only $(0,q)$ forms are $(0,0)$ forms and $(0,2)$
forms both of which are neutral under $g_N$ and $g_M$,
while on $T^4$ we also have a pair of $(0,1)$ forms $dz_1$ and
$dz_2$ which we have chosen to carry $(g_N, g_M)$
eigenvalues $(e^{\pm 2\pi i/N}, e^{\pm 2\pi i/M})$. This gives
\ben \label{esh5}
c_1^{(0,s;0,s')}(-1) &=& {2\over MN} \quad \hbox{for $\MM=K3$}
\, ,
\nonumber \\
&=& {1\over MN} \left(2 - e^{2\pi i s/N} e^{2\pi i s'/M}
- e^{-2\pi i s/N} e^{-2\pi i s'/M}
\right) \quad \hbox{for $\MM=T^4$}\, .
\een
Using this we can express the total contribution to the
partition function from the overall motion of the D1-D5 system in
the Taub-NUT space, given by \refb{ere1} for $\MM=K3$
and the 
product of \refb{ere1} and \refb{etw1} for $\MM=T^4$,
as
\ben \label{esg1}
Z_{CM} &=& - \, e^{-2\pi i v} \, \prod_{l=1}^\infty 
\prod_{r=0}^{N-1}
(1-e^{2\pi i r/N}\,
e^{2\pi i l\rho})^{2\sum_{s=0}^{N-1}\sum_{s'=0}^{M-1}
e^{-2\pi i ls'/M} e^{-2\pi i rs/N} c_1^{(0,s;0,s')}(-1)}
\nonumber \\
&&\prod_{l=1}^\infty 
\prod_{r=0}^{N-1}
(1-e^{2\pi i r/N}\,
e^{2\pi i l\rho+2\pi i v})^{-\sum_{s=0}^{N-1}\sum_{s'=0}^{M-1}
e^{-2\pi i ls'/M} e^{-2\pi i rs/N} c_1^{(0,s;0,s')}(-1)}
\nonumber \\
&&\prod_{l=0}^\infty 
\prod_{r=0}^{N-1}
(1-e^{2\pi i r/N}\,
e^{2\pi i l\rho-2\pi i v})^{-\sum_{s=0}^{N-1}\sum_{s'=0}^{M-1}
e^{-2\pi i ls'/M} e^{-2\pi i rs/N} c_1^{(0,s;0,s')}(-1)}\, .
\een
Note that the $(1-e^{-2\pi i v})^{-2}$ factor has been absorbed into the
$l=0$ term in the last term.

Finally let us turn to the contribution to the partition function from
the motion of the D1-branes along the D5-branes. First we
consider a
single
D1-brane wrapped $w$ times along $S^1$, and count the
number of states $n(w,j,l;r,k')$ of the system carrying left-moving
momentum $l/M$  along $S^1$, $g_M$ eigenvalue
$e^{2\pi i k'/M}$, $g_N$ eigenvalue $e^{2\pi i r/N}$
and $\wt S^1$ momentum $j$. Since the 
boundary condition on various fields are twisted by $g_M$ under
$2\pi$ translation along $S^1$,
the CFT on a D1-brane wrapped $w$ times along $S^1$
satisfies boundary condition twisted by $(g_M)^w$. Furthermore
since the effective length of the D1-brane is now $2\pi w$, a
momentum $l/M$ along $S^1$ will appear as $lw/M$ units
of momentum in the CFT living on the D1-brane. 
It now follows from \refb{ek8} that\cite{0609109,0708.1270} 
\ben \label{enu1}
n(w,j,l;r,k') &=& \sum_{s=0}^{N-1}\sum_{s'=0}^{M-1}
e^{-2\pi i rs/N} e^{-2\pi i k's'/M} c_b^{(0,s;r',s')}(
4lw/M - j^2)\, ,
\nonumber \\
&&  b=\hbox{$j$ mod 2}, \quad r'=\hbox{$w$ mod $M$}\, .
\een
The requirement that 
we only keep the modes which are invariant under the
transformation $g_M$ accompanied by
$2\pi$ translation along $S^1$ forces the constraint
$k'=l$ mod $M$. It is now straightforward to evaluate the
contribution to the $g_N$ twisted
partition function from multiple states of
this type, carrying different $(w,l,j)$\cite{9608096}:
\be \label{erel}
Z_{D1D5} =e^{-2\pi i \wt\gamma\sigma}\,
\prod_{r=0}^{N-1} 
\prod_{b=0}^1 \prod_{w\in \zzz,l\in\zzz,j\in 2\zzz+b
\atop w>0, l\ge 0} \left( 1 - e^{2\pi i r/N} e^{2\pi i (\sigma w
/M + \rho l
+ vj)}\right)^{-n(w,j,l;r,l)}\, .
\ee
where
\be \label{egamdef}
\wt\gamma = \cases{ \hbox{${1/ M}$ 
for $\MM=K3$}\cr \hbox{0 for 
$\MM=T^4$}}\, .
\ee
The $e^{-2\pi i \wt\gamma\sigma}$ in \refb{erel}
accounts for the fact that the actual number of D1-branes
required to produce a total D1-brane charge $Q_1$
in the background of a D5-brane is given by $Q_1+1$ for 
$\MM=K3$ and $Q_1$ for $\MM=T^4$.
Multiplying \refb{ea2}, \refb{esg1} and \refb{erel} we get the
total partition function of the system:
\be \label{etot}
Z(\rho,\sigma,v) = Z_{KK} \, Z_{CM} \, Z_{D1D5}=
 -1 / \wt\Phi(\rho,\sigma, v)\, ,
\ee
where
\ben \label{ephi}
\wt\Phi(\rho,\sigma, v)
&=&  e^{2\pi i (\wt\alpha \rho
+\wt\gamma\sigma + \wt\beta v)} \,
\prod_{b=0}^1 \prod_{r=0}^{N-1}\prod_{r'=0}^{M-1}
\prod_{k\in\zzz+{r'\over M}, l\in\zzz, j\in 2\zzz+b\atop
k,l\ge 0, j<0 \, for\, k=l=0}
\left[ 1 - e^{2\pi i r/N} \, e^{2\pi i (k\sigma +l\rho + j v)}
\right]^a\nonumber \\
a &\equiv& \sum_{s=0}^{N-1} \sum_{s'=0}^{M-1} e^{-2 
\pi i (s'l / M +
rs/N)} c_b^{(0,s;r',s')}(4kl  - j^2)\, ,
\een
with $\wt\alpha$, $\wt\beta$, $\wt\gamma$ defined in
\refb{ek9}. Note that the $k=0$ term in this product gives
the result for $Z_{KK} Z_{CM}$.

\sectiono{Asymptotic Growth} \label{sasymp}

We now study the growth of the index for large $Q^2$, $P^2$ and
$Q.P$. This can be done by standard procedure described in
\cite{9607026,0412287,0510147,0605210}. We deform the 
three dimensional contour of integration
over $(\rho,\sigma,v)$ to small imaginary values of $(\rho,\sigma,v)$.
During this deformation we pick up contribution 
from the residues at various poles, given by the zeroes of $\wt\Phi$,
which give the leading contribution to the index, -- the contribution
from the 
final contour can be shown to be subleading 
compared to the contribution from the residues at the 
poles\cite{0708.1270}.
Thus we need to first determine the location of the zeroes of
$\wt\Phi$. This has been done in appendix \ref{szero} where
it is shown that
$\wt\Phi$ has double zeroes on the subspaces:
\be \label{ediv1a}
 n_2(\rho\sigma-v^2) -m_1\rho + n_1 \sigma + m_2 + jv = 0\, ,
\ee
for values of 
$(m_1,n_1,m_2,n_2,j)$ satisfying
\ben \label{esh3a}
&& m_1 n_1 +m_2 n_2 +{j^2\over 4} = {1\over 4}\, , 
\nonumber \\
&& m_1\in M\ZZZ, \quad m_2\in \ZZZ, \quad n_2 \in 
N\ZZZ, \quad n_1\in\ZZZ, \quad
j\in 2\ZZZ+1\, .
\een
Now the analysis of \cite{9607026,0412287,0510147,0605210} 
tells us that for large $Q^2$, $P^2$,
$Q.P$ the contribution from the residue at the pole
\refb{ediv1a} of $1/\wt\Phi$ grows as
\be \label{egrow}
\exp\left(\pi\sqrt{Q^2P^2-(Q.P)^2}/|n_2|\right) \qquad
\hbox{for $ |n_2|>0$}\, .
\ee
On the other hand the poles at $n_2=0$ are responsible for wall
crossing and their contribution grows much slower than
\refb{egrow}\cite{0702141,0702150,0705.3874,0706.2363}.
Thus the leading contribution comes from the pole at
\refb{ediv1a} for
the minimum non-zero value of $|n_2|$.
Eq.\refb{esh3a} shows that this is $N$.
Thus the index grows as
\be \label{eindgr}
\exp\left(\pi\sqrt{Q^2P^2-(Q.P)^2}/N\right)\, .
\ee
Since for this charge the black hole entropy $S_{BH}$
is given by $\pi\sqrt{Q^2P^2-(Q.P)^2}$\cite{9507090,9512031}, 
\refb{eindgr} is
in precise agreement with the macroscopic prediction
\refb{emacpred}.

\sectiono{Conclusion} \label{scon}

It is widely believed that since string theory provides us with
a consistent quantum theory of gravity, black holes in string theory
do not lead to a loss of information. If so, the black hole 
must represent an ensemble of microstates and the black hole
entropy
must have an interpretation as the logarithm of the
degeneracy of microstates. Furthermore quantum string theory around
a black hole background must contain all possible information about the
microstates. It is therefore important to learn how we can extract
information about the black hole microstates by studying quantum string
theory around the black hole background.

The results of \cite{0911.1563} and this paper provide a small step
in this direction. In these papers we discuss how to extract information
about one specific feature of the black hole microstates, namely
distribution of the $\ZZZ_N$ charges among the microstates.
Quantum string theory around the near horizon background leads to
a specific algorithm for extracting this information. Our analysis 
shows that in the limit of large charges
the results of the macroscopic analysis are in exact agreement
with the microscopic results in a wide class of models where the
latter is computable.
While using the rules of $AdS_2/CFT_1$ correspondence 
we can in principle compute
the ensemble average of more general operators on the black hole
side, in the absence of non-renormalization results it is not clear how
we might compare this with the microscopic results.

\medskip

\noindent {\bf Acknowledgment:}  
I wish to thank Nabamita Banerjee, Atish Dabholkar, Joao Gomes
and Sameer Murthy for useful discussions.
This work was supported
in part by the JC Bose fellowship of the Department of Science and
Technology, India and by the Blaise Pascal Chair, France.

\medskip

\noindent {\bf Note added:} 
I have been informed by Suresh Govandarajan that the
modular forms of subgroups of $Sp(2,\ZZZ)$ which appear here have
also been constructed independently in \cite{suresh1,suresh2}.

\appendix

\sectiono{Explicit computation of $c_b^{(r,s;r',s')}$} \label{sb}

In this appendix we shall describe the strategy for explicit
computation of the right hand side of \refb{ek8}
\ben \label{ek8aa}
F^{(r,s;r',s')}(\tau,z) &\equiv&
{1\over MN} Tr_{RR;g_M^{r'} g_N^r} \left (g_M^{s'} g_N^{s} 
(-1)^{J_L+J_R} e^{2\pi i (\tau L_0-\bar\tau \bar L_0)}
e^{2\pi i J_L z} \right)\nonumber \\
&=& \sum_{b=0}^1 \sum_{j\in 2\zzz+b, n\in \zzz/MN} 
c_b^{(r,s;r',s')}
(4n - j^2) e^{2\pi i (n\tau + j z)}\, ,
\een
and hence of $c_b^{(r,s;r',s')}$. First of all $F^{(0,0;0,0)}$
is simply $1/MN$ times the elliptic genus of $\MM$ and is
given by\cite{9306096}
\ben \label{ef00}
F^{(0,0;0,0)}(\tau,z) &=& 0 \quad \hbox{for $\MM=T^4$}\nonumber
\\
&=& {8\over MN} \left[ {\vt_2(\tau,z)^2\over \vt_2(\tau,0)^2}
  +  {\vt_3(\tau,z)^2\over \vt_3(\tau,0)^2}
 +  {\vt_4(\tau,z)^2\over \vt_4(\tau,0)^2}
\right]\quad \hbox{for $\MM=K3$}\, ,
\een
where $\vt_i$ are the Jacobi theta functions.
For 
non-vanishing $r$
and/or $r'$ the contribution to the right hand side of \refb{ek8aa},
coming from twisted sectors localized at the fixed points of
$g_M^{r'} g_N^r$, can be computed by taking the size of K3 to
be large so that the
geometry near the fixed points is nearly
flat. In this case the $\sigma$ model near the fixed point
can be regarded as the orbifold of a free field theory, and the
action of $g_N^r g_M^{r'}$ near the fixed point may be
represented by a rotation by some angle $2\pi\theta$ in one plane
and rotation by an opposite angle $-2\pi\theta$ in the orthogonal plane.
The action of $g_M^{s'} g_N^{s}$ near such a fixed point can
be of two types, -- either it takes the fixed point to a different
fixed point or it leaves the fixed point fixed. In the former case
the contribution to the trace in  \refb{ek8aa} vanishes, whereas
in the latter case the action of $g_M^{s'} g_N^{s}$ near 
the fixed point can be represented by a rotation by $2\pi\phi$
in one plane and a rotation by $-2\pi\phi$ in the orthogonal plane. This
gives 
a contribution to \refb{ek8aa}  of the form
\ben \label{efixed1}
&& {1\over MN} \prod_{n=1}^\infty \bigg\{ \left(1 - q^{n+\theta-1}
e^{2\pi i \phi}\right)^{-2}\left(1 - q^{n-\theta }
e^{-2\pi i \phi}\right)^{-2} \left(1 - q^{n+\theta-1}
e^{2\pi i \phi} e^{2\pi i z}\right)\nonumber \\
&& \qquad \qquad \qquad \left(1 - q^{n+\theta-1}
e^{2\pi i \phi} e^{-2\pi i z}\right)
\left(1 - q^{n-\theta}
e^{-2\pi i \phi} e^{2\pi i z}\right)\left(1 - q^{n-\theta}
e^{-2\pi i \phi} e^{-2\pi i z}\right)\nonumber \\
&=& {1\over MN} {\vt_1 (\tau, z+\theta\tau +\phi)
\vt_1 (\tau, -z+\theta\tau +\phi)\over \vt_1(\tau, \theta\tau+\phi)^2}\, ,
\een
where $q\equiv e^{2\pi i \tau}$.
The full contribution to $F^{(r,s;r',s')}(\tau,z)$ is obtained by
summing over the contribution from all the fixed points of
$g_N^r g_M^{r'}$ which are also fixed by $g_M^{s'} g_N^{s}$.
Finally $F^{(0,s;0,s')}$ can be computed using the modular
transformation rules\cite{9306096}
\be \label{emodular}
F^{(r,s;r',s')}\left( {a\tau+b\over c\tau+d}, {z\over c\tau+d}
\right) = \exp\left[2\pi i {cz^2\over c\tau+d}\right]
F^{(cs+ar,ds+br;cs'+ar',ds'+br')}(\tau,z)\, .
\ee

This gives a way to compute $F^{(r,s;r',s')}(\tau,z)$ using purely
geometric data, namely the fixed points of the different elements of
$\ZZZ_M\times \ZZZ_N$ and the action of the elements of
$\ZZZ_M\times \ZZZ_N$ near the fixed points.
Note that
$F^{(r,s;r',s')}(\tau,z)$ constructed from
\refb{efixed1}, \refb{emodular} is invariant under
$z\to -z$. This is a consequence of the $SU(2)_L$ 
R-symmetry of
the underlying conformal field theory that allows us to change
the sign of $J_L$.

\sectiono{Threshold Integral Representation of $\wt\Phi$}
\label{sthreshold}

In this appendix we shall describe a threshold integral representation
of $\wt\Phi$.
For this we define
\be\label{ehbexp}
h_b^{(r,s;r',s')}(\tau) = \sum_{k\in{1\over MN}\zzz -{b^2\over 4}}
c_b^{(r,s;r',s')}(4k) e^{2\pi i k\tau}, 
\ee
\be\label{edefomega}
\Omega=\pmatrix{\rho  & v \cr v  & \sigma}\, ,
\ee
and
\bea{e7n}
{1\over 2} p_R^2 &=& {1\over 4 \det Im  \Omega} |-m_1 \rho  +
m_2 + n_1 \sigma + n_2 (\sigma\rho -v^2) + j v |^2, \nonumber \\
{1\over 2} p_L^2
&=& {1\over 2}  p_R^2 + m_1 n_1 + m_2 n_2 + {1\over 4} j^2\, .
\eea
We now consider the `threshold integral'
\be\label{rwthrint}
\tI(\rho , \sigma, v ) = \sum_ {r, s =0}^{N-1}
\sum_{r',s'=0}^{M-1}
\sum_{b=0}^1
\tI_{r, s;r',s'; b}\, , \ee
where
\ben\label{defirsl}
\tI_{r,s;r',s';b} &=& \int_{\FF} \frac{d^2\tau}{\tau_2} 
\bigg[\sum_{\stackrel{m_1\in\zzz, m_2\in \zzz/N, n_2 \in 
N\zzz -r}{n_1\in \zzz+ \frac{r'}{M}, 
j\in 2\zzz + b}}
q^{p_L^2/2} \bar q^{ p_R^2/2} e^{2\pi i  m_1 s'/M}
e^{-2\pi i m_2 s}
h_b^{(r,s;r',s')}(\tau) \nonumber \\
&& - \delta_{b,0} \delta_{r,0} \delta_{r',0}
c^{(0,s;0,s')}_0(0)
\bigg]\, ,
\een
with
\be\label{edefq}
q\equiv e^{2\pi i\tau}\, .
\ee
$\FF$ denotes the fundamental region of $SL(2,\ZZZ)$ in the
upper half plane. The subtraction terms 
proportional to $c_0^{(0,s;0,s')}(0)$ have been chosen so that the
integrand vanishes faster than $1/\tau_2$
in the $\tau\to i\infty$ limit, rendering the
integral finite. 

Following the manipulations outlined in
\cite{0602254} following earlier work of
\cite{dixon,9512046,9607029} one can show that
\be \label{esh1}
\wt \II (\rho,\sigma, v) = -2\ln [ (\det Im~\Omega)^{\wt k} ]
- 2\ln \wt\Phi(\rho,\sigma, v) - 2\ln \check\Phi(\bar\rho,\bar\sigma,\bar v)
+\hbox{constant}
\ee
where $\wt\Phi$ has been defined in \refb{ek7} 
and\footnote{Since the exponent $a$ in \refb{ek7}, \refb{ek7check}
is a real number (in fact an integer) we have
$\check\Phi(\bar\rho,\bar \sigma, \bar v) = \overline{\wt\Phi(\rho,
\sigma,v)}$.}
\ben \label{ek7check}
\check\Phi(\bar\rho,\bar\sigma,\bar v) &=&  e^{-2\pi i (\wt\alpha 
\bar\rho
+\wt\gamma\bar\sigma + \wt\beta \bar v)} \,
\prod_{b=0}^1 \prod_{r=0}^{N-1}\prod_{r'=0}^{M-1}
\prod_{k\in\zzz+{r'\over M}, l\in\zzz, j\in 2\zzz+b\atop
k,l\ge 0, j<0 \, for\, k=l=0}
\left[ 1 - e^{-2\pi i r/N} \, e^{-2\pi i (k\bar\sigma +l\bar\rho + j \bar v)}
\right]^a\nonumber \\
a &\equiv& \sum_{s=0}^{N-1} \sum_{s'=0}^{M-1} e^{-2 
\pi i (s'l / M +
rs/N)} c_b^{(0,s;r',s')}(4kl - j^2)\, ,
\een
\ben \label{esh2}
\wt k &=& {1\over 2} \sum_{s=0}^{N-1}\sum_{s'=0}^{M-1} 
c_0^{(0,s;0,s')}(0)
\, .
\een
Since the detailed analysis of a specific integral of this type has been
carried out in \cite{0602254}, we shall only
describe the basic steps
leading to \refb{esh1}, focussing on the main differences between
the general case and the special case analyzed in \cite{0602254}.
\begin{enumerate}
\item We first carry out Poisson resummation over $m_1$ and $m_2$
to express $\wt\II_{r,s;r',s';b}$ as
\be \label{etry1}
\int_\FF {d^2\tau\over \tau_2^2} \, {Y\over \rho_2}\, 
\bigg[ N\sum_{{k_1\in \zzz+{s'\over M}, k_2\in N\zzz-s\atop
n_1\in \zzz+{r'\over M}, n_2\in N\zzz-r, j\in2\zzz+b}}
h_b^{(r,s;r',s')}(\tau)  
e^{\GG(\vec n,
\vec k, j)} -  \delta_{b,0} \delta_{r,0} \delta_{r',0}
c_0^{(0,s;0,s')}(0)
\bigg]
\ee
where
\ben\label{fing}
{\cal G}(\vec n, \vec k, j) &=& 
- 
\frac{\pi Y}{\rho_2^2 \tau_2}|{\cal A}|^2 
- 2\pi i \sigma {\rm det}\,  A 
+\frac{\pi j}{\rho_2} ( v\tilde {\cal A} - \bar v {\cal A})
- \frac{\pi n_2}{\rho_2} ( v^2 \tilde{\cal A} - \bar v^2 {\cal A})
\cr && +
\frac{2 \pi i v_2^2}{\rho_2^2} ( n_1 + n_2 \bar \rho) {\cal A}
 + 2\pi i \tau \frac{j^2}{4}\, ,
\een
\be\label{fing1}
Y\equiv
\rho_2\sigma_2 - v_2^2\, , \qquad A \equiv \left(
\begin{array}{cc}
n_1 & k_1 \\
n_2 & k_2 
\end{array}
\right)\, , \quad 
{\cal A} \equiv ( 1 , \rho) A 
\left(
\begin{array}{c}
\tau \\ 1
\end{array}
\right)\, ,
\quad
\tilde {\cal A} \equiv ( 1 , \bar \rho) A 
\left(
\begin{array}{c}
\tau \\ 1
\end{array}
\right)\, .
\ee

\item Using  \refb{rwthrint}, \refb{etry1} and performing the sum over
$j$ using \refb{ek8aa}, \refb{ehbexp}
we get
\be \label{etry3}
\wt\II = \int_\FF {d^2\tau\over \tau_2^2}\, 
\bigg[N\sum_{k_1, n_1\in{\zzz\over M}, k_2,n_2\in\zzz}\JJ(A,\tau)
- \sum_{s=0}^{N-1}\sum_{s'=0}^{M-1} c_0^{(0,s;0,s')}(0)\bigg]\, ,
\ee
where
\ben \label{etry4}
\JJ(A,\tau) &=& \frac{Y}{\rho_2}
\, 
\exp\Bigg(  - \frac{\pi Y}{\rho_2^2 \tau_2}|{\cal A}|^2
- 2\pi i \sigma {\rm det}\,  A \nonumber \\
&& - \frac{\pi n_2}{\rho_2} ( v^2
\tilde{\cal A} -
\bar v^2 {\cal A}) + \frac{2 \pi i v_2^2}{\rho_2^2} 
( n_1 + n_2 \bar \rho) {\cal
A} \Bigg) \, F^{(r,s;r',s')}\left(\tau, -i { v\tilde {\cal A} -
\bar v
{\cal A} \over
2\, \rho_2}\right) \nonumber \\
&&    r =-n_2 \, \hbox{mod $N$}, \, s =  
-k_2 \, \hbox{mod $N$}, \, r' =
M n_1 \, \hbox{mod $M$}, \, s' =
M k_1 \, \hbox{mod $M$}\, . \nonumber \\
\een

\item Using \refb{emodular} one can show that 
\be \label{emodsym}
\JJ\left(A \pmatrix{a & b\cr c & d},\tau\right)
= \JJ\left(A,  {a\tau+b
\over c\tau+d}\right), 
\qquad \pmatrix{a &b\cr c & d}\in SL(2,\ZZZ)\, .
\ee
Under the map $\tau\to (a\tau+b)/(c\tau+d)$
the fundamental region $\FF$ of $SL(2,\ZZZ)$
gets mapped to its image.
With the help of \refb{emodsym} we can restrict
the sum over the matrix $A$ to the following ranges\cite{dixon}:
\begin{enumerate}
\item Non-degenerate orbit:
\be\label{enondeg}
A=\pmatrix{k & m\cr 0 & p}, \quad k,m\in {\ZZZ\over M}, \quad
p\in\ZZZ, \quad p\ne 0, \quad 0\le m<k\, ,
\ee
with the integration over $\tau$ ranging over two copies of the upper
half plane.
\item Degenerate orbit:
\be \label{edeg}
A=\pmatrix{0 & m\cr 0 & p}, \quad m\in {\ZZZ\over M}, \quad p\in
\ZZZ, \quad (m,p)\ne (0,0)\, ,
\ee
with the integration over $\tau$ ranging over the strip $-{1\over 2}\le
\tau_1\le {1\over 2}$, $\tau_2>0$.
\item Zero orbit: 
\be \label{ezorbit}
A=\pmatrix{0 & 0\cr 0 & 0}\, ,
\ee
with the integration over $\tau$ ranging over the fundamental domain
$\FF$.
\end{enumerate}
We shall briefly discuss the computation of the contribution from the
non-degenerate orbits. The analysis of the contribution from the other
orbits follows \cite{0602254}.

\item
The steps needed for evaluating the contribution from the non-degenerate
orbits are as follows:
\begin{enumerate}
\item We first expand $F^{(r,s;r',s')}(\tau,z)$
using \refb{ek8aa} and change integration variable from $\tau_1$ to 
$\tau_1'$:
\be \label{edeftau}
\tau_1'=\tau_1+{m\over k} +{p\over k}\rho_1\, .
\ee
Since for non-degenerate orbit the $\tau_1$ integration ranges from
$-\infty$ to $\infty$, $\tau_1'$ integral will also range from
$-\infty$ to $\infty$.
\item After this change of variables the $m$ dependence of the integral
becomes simple.
The explicit dependence on $m$ takes the form 
$e^{-2\pi i m n/k}$, $n$ and $j$ being the
integers appearing in the expansion \refb{ek8aa}.
There is also an implicit dependence on $m$
through $c_b^{(r,s;r',s')}
(4n-j^2)$ since $s'=Mm$ mod $M$ due to \refb{etry4}.
The
sum over $m$ can be performed by
expanding the summation range of $m$ to $0\le m< Mk$ in steps
of $1/M$, at the
cost of having to divide the sum by a factor of $M$.
We then
sum over all values of $m$ of the form ${s'\over M} +\hbox{integer}$
for fixed $s'$ and finally sum over integral $s'$ in the range
$0\le s'\le (M-1)$. 
For fixed $s'$ the sum takes the form
\be \label{eintsum}
 \sum_{m\in \zzz + {s'\over M}, 0\le m<Mk-1}
\, e^{-2\pi i m n/k} = Mk\,
e^{-2\pi i n s'/(M k)}\, \sum_{l\in\zzz} \delta_{n,kl}
= Mk\, \sum_{l\in\zzz} e^{-2\pi i ls'/M}\, \delta_{n,kl}\, .
\ee
The sum over $n$ can now be performed using the Kronecker delta.
\item Next we carry out the $\tau_1'$ integral which is Gaussian and the
$\tau_2$ integral using the identity
\be \label{eidentity}
\int_0^\infty {du\over u^{3/2}} e^{-au-bu^{-1}}
= \sqrt{\pi\over b} e^{-2\sqrt{ab}}\, .
\ee
At this stage the $p$ dependent part of the summand takes the form
\be \label{epdep}
{1\over |p|}\exp\left\{  
  -2\pi i \sigma kp - 2\pi k|p| \sigma_2 - 2\pi kp \sigma_2 
   - 2\pi i  lp\rho_1 - 2\pi l|p| \rho_2 
   -2\pi i j p v_1 - 2\pi j |p| v_2 \right\}\, .
\ee
\item Finally we perform the sum over $p$ by breaking it  into
contribution from $p>0$ and $p<0$ terms, 
making a change of variables $p\to -p$ in the $p<0$
terms, and using the identity
\ben \label{epsum}
\sum_{p\in N\zzz+s,p>0} \, {1\over p}\, e^{2\pi i \alpha p}
&=& {1\over N} \sum_{r=0}^{N-1} \sum_{p\in \zzz,p>0} 
e^{2\pi i r (p-s)/N}
\, {1\over p}\, e^{2\pi i \alpha p} \nonumber \\
&=&
- {1\over N} \sum_{r=0}^{N-1} e^{-2\pi i rs/N}
\, \ln (1 - e^{2\pi i r/N} e^{2\pi i \alpha})
\, .
\een
\end{enumerate}
\item
At the end of this manipulation we get the contribution from the
non-degenerate orbits to be
\ben \label{efinnondeg}
&& - 2 \sum_{r,s=0}^{N-1} \sum_{r',s'=0}^{M-1}\sum_{b=0}^1
\sum_{k\in \zzz+{r'\over M}, l\in\zzz, j\in 2\zzz+b, k>0, l\ge 0}
e^{-2\pi i rs/N} e^{-2\pi i ls'/M} c_b^{(0,s;r',s')}(4kl-j^2)
\nonumber \\ && \qquad 
\left\{
\ln\left[ 1 - e^{2\pi i r/N} e^{2\pi i(k\sigma + l\rho+jv)}\right]
+  \ln\left[ 1 - e^{-2\pi i r/N} e^{-2\pi i(k\bar\sigma + l\bar\rho
+j\bar v)}\right]
\right\} +\hbox{constant}\, .\nonumber \\
\een
\end{enumerate}
One can recognize \refb{efinnondeg} as the contribution from
$-2\ln\wt\Phi(\rho,\sigma,v) - 2\ln \check\Phi(\bar\rho,
\bar\sigma,\bar v)$ 
except for the $k=0$ terms and the overall
multiplicative factors in the product representation
\refb{ek7} of $\wt\Phi$ and \refb{ek7check}
of $\check\Phi$. By carefully analyzing the contribution
from the degenerate and the zero orbits one recovers the 
complete set of terms on the left hand side of
\refb{esh1}. 

\sectiono{Zeroes of $\wt\Phi$} \label{szero}

In this appendix we shall determine the locations of the zeroes
of the function $\wt\Phi$. 
{}From \refb{esh1} it follows that at the locations of the zeroes and
poles of $\wt\Phi$ we have logarithmic divergences in
$\wt \II$.  Thus we can determine the locations of the
zeroes and poles of $\wt\Phi$ by determining the singularities
of $\wt\II$. 
Since the integrand in $\wt\II$ is finite for finite $\tau$, and
in its original form given in \refb{rwthrint}, \refb{defirsl}
the integration over $\tau$
runs over the fundamental
domain of $SL(2,\ZZZ)$, the only possible source of divergence
is from the region of large $\tau_2$. This requires that the
powers of $q$($\equiv e^{2\pi i\tau}$ ) and $\bar q$ be equal
so that the $\tau_1$ integral for large $\tau_2$ does not vanish,
and non-positive so that the $\tau_2$ integral diverges. Since the
only dependence on $\bar q$ is through the $\bar q^{p_R^2/2}$
term, and since $p_R^2$ is positive semi-definite, this requires
$p_R^2$ to vanish. Furthermore $p_L^2$ is also
positive semidefinite
(although it is not directly apparent from its
definition), and hence the only possible way to produce a divergence
is to get a non-positive power of $q$ from the expansion of
$h_b^{(r,s;r',s')}(\tau)$.\footnote{There is a further restriction coming
from the fact that for $p_L^2=p_R^2=0$ all the $m_i$'s, $n_i$'s and
$j$ must vanish. The divergence in the $\tau_2$ integral from such a term is
removed by the subtraction term in \refb{defirsl}. 
Thus a divergent $\tau_2$
integral requires $p_L^2$ to be strictly positive. This explains the
strict inequality $\beta< {1\over 4}$ in \refb{esh3}.}  
Such terms are quite restricted, and one
finds that the possible divergences in $\wt\II$ arise 
at\cite{0605210,0708.1270}
\be \label{ediv1}
 n_2(\rho\sigma-v^2) -m_1\rho + n_1 \sigma + m_2 + jv = 0\, ,
\ee
for values of 
$(m_1,n_1,m_2,n_2,j)$ satisfying
\ben \label{esh3}
&& m_1 n_1 +m_2 n_2 +{j^2\over 4} = {1\over 4}-\beta, \qquad
0\le\beta <{1\over 4}\, , 
\nonumber \\
&& m_1\in\ZZZ, \quad m_2\in \ZZZ/N, \quad n_2 \in 
\ZZZ, \quad n_1\in\ZZZ/M, \quad
j\in 2\ZZZ+1\, .
\een
At this point $p_R^2$ vanishes and $p_L^2/2 = (1/4)-\beta$ so that 
the $q^{p_L^2/2} \bar q^{p_R^2/2}$ factor becomes purely a
function of $\tau$ and not of $\bar\tau$. 
This has to be cancelled against a similar
$\tau$ dependent factor in $h_b$ so that the $\tau_1$ integral is
finite and we get a logarithmically divergent $\tau_2$ 
integral.
The coefficient of the divergent term can be easily determined from
\refb{defirsl}, \refb{ehbexp} and \refb{esh1} 
and we get, near this point,\footnote{We need to
account for the fact that the terms in the expression for
$\wt\II$ with $(m_i,n_i,j)$ and $(-m_i, -n_i, -j)$ give
identical results.}
\ben \label{esh4}
\wt \Phi (\rho,\sigma, v) &\sim&
\left(n_2(\rho\sigma-v^2) -m_1\rho + n_1 \sigma + m_2 + jv
\right)^{\sum_{s'=0}^{M-1}\sum_{s=0}^{N-1}
e^{2\pi i m_1 s'/M} e^{-2\pi i m_2 s}  c^{(r,s;r',s')}_1(-1+4\beta)}\, ,
\nonumber \\
&& r=\hbox{$-n_2$ mod $N$}, \quad \hbox{$r'=Mn_1$ mod
$M$}\, .
\een
Let us first focus on the $r=r'=0$ terms, \i.e.\
configurations with $n_2\in N\ZZZ$,
$n_1\in\ZZZ$. In this case it follows from
\refb{esh3} that $\beta$ must vanish and so the right hand side
of \refb{esh4} involves $c_1^{(0,s;0,s')}(-1)$. 
Using \refb{esh5} we now get,
for $\MM=K3$,
\be \label{esh6}
\sum_{s'=0}^{M-1}\sum_{s=0}^{N-1}
e^{2\pi i m_1 s'/M} e^{-2\pi i m_2 s}  c^{(0,s;0,s')}_1(-1)
=\cases{ \hbox{2 for $m_1\in M\ZZZ$, $m_2\in \ZZZ$}
\cr
\hbox{0 otherwise}\, ,
}
\ee
and for $\MM=T^4$,
\be \label{esh7}
\sum_{s'=0}^{M-1}\sum_{s=0}^{N-1}
e^{2\pi i m_1 s'/M} e^{-2\pi i m_2 s}  c^{(0,s;0,s')}_1(-1)
=\cases{ \hbox{2 for $m_1\in M\ZZZ$, $m_2\in \ZZZ$}
\cr \hbox{$-1$ for $m_1\in M\ZZZ\pm 1$, $m_2\in \ZZZ
\mp {1\over N}$}\cr
\hbox{0 otherwise}\, .
}
\ee
In either case the exponent is positive, producing zeroes
of $\wt\Phi$, only for $m_1\in M\ZZZ$ and $m_2\in \ZZZ$.
Thus the net result is that the only zeroes of $\wt\Phi$ for $r=r'=0$
are
of the form:
\ben \label{esh8}
\wt \Phi (\rho,\sigma, v) &\sim&
\left(n_2(\rho\sigma-v^2) -m_1\rho + n_1 \sigma + m_2 + jv
\right)^{2}\, ,
\nonumber \\
&& m_1\in M\ZZZ, \quad m_2\in \ZZZ,
\quad n_1\in\ZZZ, \quad n_2\in N\ZZZ\, .
\een
Let us now analyze the contribution to the exponent in \refb{esh4}
when $r$ and/or $r'$ is non-zero. 
{}From the definition of the coefficients
$c_b^{(r,s;r',s')}$ given in \refb{ek8} it follows that the
exponent in \refb{esh4} can be interpreted as the
number of states weighted by $(-1)^{J_L+J_R}$
in the sector twisted by $g_N^r g_M^{r'}$,
and carrying $g_M$ eigenvalue $e^{-2\pi i m_1/M}$,
$g_N$ eigenvalue $e^{2\pi i m_2}$, $J_L=\pm 1$,
$L_0=\beta$ and $\bar L_0=0$. If $g_M^{r'}g_N^r$ does not
have a fixed point in $\MM$ then this number is zero since by
taking the size of $\MM$ to be sufficiently large we can ensure that
there will be no twisted sector state with $\bar L_0=0$.
If $g_M^{r'}g_N^r$ has fixed points in $\MM$, then
we can compute this number by taking the size
of $\MM$ to be large so that near the fixed points we can
regard the space as almost flat, with $g_N^rg_M^{r'}$ acting as
 rotation by some angle $\theta$ in one plane and by $-\theta$ 
 in an orthogonal plane. Under such rotations all the bosons
 and fermions in the sigma model with target space $\MM$ are
 twisted and hence there are no zero modes. Thus we have a
 unique ground state with $(J_L=J_R=0)$. Even after we apply
 left-moving oscillators to create states with $L_0=\beta$,
 $J_L=\pm 1$,  the
 states will continue to have $J_R=0$. 
 Thus the weight factor $(-1)^{J_L+J_R}$ is always $-1$ for
 $J_L=\pm 1$ and
 hence the exponent of \refb{esh4} is always negative. This shows
 that \refb{esh4} never gives a zero of $\wt\Phi$ for $r$
 and/or $r'$ non-zero, and
 the only zeroes of $\wt\Phi$ are of the form given in \refb{esh8}.



\begin{thebibliography}{99}


\bibitem{9505054}
  S.~Chaudhuri, G.~Hockney and J.~D.~Lykken,
  ``Maximally Supersymmetric String Theories In D $<$ 10,''
  Phys.\ Rev.\ Lett.\  {\bf 75}, 2264 (1995)
  [arXiv:hep-th/9505054].

\bibitem{9506048}
  S.~Chaudhuri and J.~Polchinski,
  ``Moduli space of CHL strings,''
  Phys.\ Rev.\  D {\bf 52}, 7168 (1995)
  [arXiv:hep-th/9506048].

\bibitem{0510147}
  D.~P.~Jatkar and A.~Sen,
  ``Dyon spectrum in CHL models,''
  JHEP {\bf 0604}, 018 (2006)
  [arXiv:hep-th/0510147].

\bibitem{0602254}
  J.~R.~David, D.~P.~Jatkar and A.~Sen,
  ``Product representation of dyon partition function in CHL models,''
  JHEP {\bf 0606}, 064 (2006)
  [arXiv:hep-th/0602254].
  
\bibitem{0603066}
  A.~Dabholkar and S.~Nampuri,  
  ``Spectrum of dyons and black holes in 
  CHL orbifolds using Borcherds lift,''
  arXiv:hep-th/0603066.

\bibitem{0605210}
  J.~R.~David and A.~Sen,
  ``CHL dyons and statistical entropy function from D1-D5 system,''
  JHEP {\bf 0611}, 072 (2006)
  [arXiv:hep-th/0605210].

\bibitem{0607155}
  J.~R.~David, D.~P.~Jatkar and A.~Sen,
  ``Dyon spectrum in N = 4 supersymmetric type II string theories,''
  arXiv:hep-th/0607155.


\bibitem{0609109}
  J.~R.~David, D.~P.~Jatkar and A.~Sen,
  ``Dyon spectrum in generic N = 4 supersymmetric Z(N) orbifolds,''
  arXiv:hep-th/0609109.

\bibitem{0612011}
  A.~Dabholkar and D.~Gaiotto,
  ``Spectrum of CHL dyons from genus-two partition function,''
  arXiv:hep-th/0612011.

\bibitem{9507090}
  M.~Cvetic and D.~Youm,
  ``Dyonic BPS saturated black holes of heterotic string on a six torus,''
  Phys.\ Rev.\  D {\bf 53}, 584 (1996)
  [arXiv:hep-th/9507090].

\bibitem{9512031}
M.~Cvetic and A.~A.~Tseytlin,
``Solitonic strings and BPS saturated dyonic black holes,''
  Phys.\ Rev.\  D {\bf 53}, 5619 (1996)
  [Erratum-ibid.\  D {\bf 55}, 3907 (1997)]
  [arXiv:hep-th/9512031].

\bibitem{9602060}
B.~de Wit,
``N = 2 electric-magnetic duality in a chiral background,''
Nucl.\ Phys.\ Proc.\ Suppl.\  {\bf 49}, 191 (1996)
[arXiv:hep-th/9602060].

\bibitem{9603191}
B.~de Wit,
``N=2 symplectic reparametrizations in a chiral background,''
Fortsch.\ Phys.\  {\bf 44}, 529 (1996)
[arXiv:hep-th/9603191].

\bibitem{9812082}
G.~Lopes Cardoso, B.~de Wit and T.~Mohaupt,
``Corrections to macroscopic supersymmetric black-hole entropy,''
Phys.\ Lett.\ B {\bf 451}, 309 (1999)
[arXiv:hep-th/9812082].

\bibitem{0007195}
T.~Mohaupt,
``Black hole entropy, special geometry and strings,''
Fortsch.\ Phys.\  {\bf 49}, 3 (2001)
[arXiv:hep-th/0007195].


\bibitem{9607026}
R.~Dijkgraaf, E.~P.~Verlinde and H.~L.~Verlinde,
``Counting dyons in N = 4 string theory,''
Nucl.\ Phys.\ B {\bf 484}, 543 (1997)
[arXiv:hep-th/9607026].

\bibitem{0412287}
  G.~Lopes Cardoso, B.~de Wit, J.~Kappeli and T.~Mohaupt,
  ``Asymptotic degeneracy of dyonic 
  N = 4 string states and black hole
  entropy,''
  JHEP {\bf 0412}, 075 (2004)
  [arXiv:hep-th/0412287].

\bibitem{0505094}
D.~Shih, A.~Strominger and X.~Yin,
``Recounting dyons in N = 4 string theory,''
arXiv:hep-th/0505094.

\bibitem{0506249}
D.~Gaiotto,
``Re-recounting dyons in N = 4 string theory,''
arXiv:hep-th/0506249.

\bibitem{0508174}
  D.~Shih and X.~Yin,
  ``Exact black hole degeneracies and the topological string,''
  JHEP {\bf 0604}, 034 (2006)
  [arXiv:hep-th/0508174].

\bibitem{0705.1433}
  N.~Banerjee, D.~P.~Jatkar and A.~Sen,
  ``Adding charges to N = 4 dyons,''
  arXiv:0705.1433 [hep-th].

\bibitem{0802.0544}
  S.~Banerjee, A.~Sen and Y.~K.~Srivastava,
  ``Generalities of Quarter BPS Dyon 
Partition Function and Dyons of Torsion
  Two,''
  arXiv:0802.0544 [hep-th].

\bibitem{0802.1556}
  S.~Banerjee, A.~Sen and Y.~K.~Srivastava,
  ``Partition Functions of Torsion $>1$ Dyons in Heterotic
String Theory on $T^6$,''
  arXiv:0802.1556 [hep-th].

\bibitem{0803.2692}
  A.~Dabholkar, J.~Gomes and S.~Murthy,
  ``Counting all dyons in N =4 string theory,''
  arXiv:0803.2692 [hep-th].


\bibitem{0809.3304}
  A.~Sen,
  ``Quantum Entropy Function from AdS(2)/CFT(1) Correspondence,''
  Int.\ J.\ Mod.\ Phys.\  A {\bf 24}, 4225 (2009)
  [arXiv:0809.3304 [hep-th]].

\bibitem{0903.1477}
  A.~Sen,
  ``Arithmetic of Quantum Entropy Function,''
  JHEP {\bf 0908}, 068 (2009)
  [arXiv:0903.1477 [hep-th]].

\bibitem{0911.1563}
  A.~Sen,
  ``A Twist in the Dyon Partition Function,''
  arXiv:0911.1563 [hep-th].

\bibitem{wil1}
  M.~G.~Alford, J.~March-Russell and F.~Wilczek,
  ``DISCRETE QUANTUM HAIR ON BLACK 
  HOLES AND THE NONABELIAN AHARONOV-BOHM
  EFFECT,''
  Nucl.\ Phys.\  B {\bf 337}, 695 (1990).

\bibitem{wil2}
  S.~R.~Coleman, J.~Preskill and F.~Wilczek,
  ``Quantum hair on black holes,''
  Nucl.\ Phys.\  B {\bf 378}, 175 (1992)
  [arXiv:hep-th/9201059].



\bibitem{9708062}
  A.~Gregori, E.~Kiritsis, C.~Kounnas, N.~A.~Obers, 
  P.~M.~Petropoulos and B.~Pioline,
  ``R**2 corrections and non-perturbative 
  dualities of N = 4 string ground
  states,''
  Nucl.\ Phys.\ B {\bf 510}, 423 (1998)
  [arXiv:hep-th/9708062].

\bibitem{9708130}
  E.~Kiritsis,
  ``Introduction to non-perturbative string theory,''
  arXiv:hep-th/9708130.

\bibitem{0810.3472}
  N.~Banerjee, D.~P.~Jatkar and A.~Sen,
  ``Asymptotic Expansion of the N=4 Dyon Degeneracy,''
  JHEP {\bf 0905}, 121 (2009)
  [arXiv:0810.3472 [hep-th]].

\bibitem{0904.4253}
  S.~Murthy and B.~Pioline,
  ``A Farey tale for N=4 dyons,''
  JHEP {\bf 0909}, 022 (2009)
  [arXiv:0904.4253 [hep-th]].

\bibitem{9508144}
S.~Chaudhuri and D.~A.~Lowe,
``Type IIA heterotic duals with maximal supersymmetry,''
Nucl.\ Phys.\ B {\bf 459}, 113 (1996)
[arXiv:hep-th/9508144].

\bibitem{9508154}
P.~S.~Aspinwall,
``Some relationships between dualities in string theory,''
Nucl.\ Phys.\ Proc.\ Suppl.\  {\bf 46}, 30 (1996)
[arXiv:hep-th/9508154].

\bibitem{0702141}
  A.~Sen,
  ``Walls of marginal stability and dyon spectrum in N = 4 supersymmetric
  string theories,''
  arXiv:hep-th/0702141.

\bibitem{0702150}
  A.~Dabholkar, D.~Gaiotto and S.~Nampuri,
  ``Comments on the spectrum of CHL dyons,''
  arXiv:hep-th/0702150.
  
\bibitem{0705.3874}
  A.~Sen,
  ``Two Centered Black Holes and N=4 Dyon Spectrum,''
  arXiv:0705.3874 [hep-th].
    

\bibitem{0706.2363}
  M.~C.~N.~Cheng and E.~Verlinde,
  ``Dying Dyons Don't Count,''
  arXiv:0706.2363 [hep-th].


\bibitem{0708.1270}
  A.~Sen,
  ``Black Hole Entropy Function, 
Attractors and Precision Counting of
  Microstates,''
  arXiv:0708.1270 [hep-th].

 \bibitem{pope}
 C.~N.~Pope,
  ``Axial Vector Anomalies And The Index Theorem In Charged 
 Schwarzschild And
  Taub - Nut Spaces,''
  Nucl.\ Phys.\ B {\bf 141}, 432 (1978).

\bibitem{9912082}
J.~P.~Gauntlett, N.~Kim, J.~Park and P.~Yi,
 ``Monopole dynamics and BPS dyons in 
N = 2 super-Yang-Mills theories,''
  Phys.\ Rev.\ D {\bf 61}, 125012 (2000)
  [arXiv:hep-th/9912082].

\bibitem{0503217}
  D.~Gaiotto, A.~Strominger and X.~Yin,
  ``New connections between 4D and 5D black holes,''
  JHEP {\bf 0602}, 024 (2006)
  [arXiv:hep-th/0503217].

\bibitem{9608096}
  R.~Dijkgraaf, G.~W.~Moore, E.~P.~Verlinde and H.~L.~Verlinde,
  ``Elliptic genera of symmetric products and second quantized strings,''
  Commun.\ Math.\ Phys.\  {\bf 185}, 197 (1997)
  [arXiv:hep-th/9608096].

\bibitem{suresh1}
S.~Govindarajan, talk at NSM 2010, IIT Bombay.

\bibitem{suresh2}
S.~Govindarajan, "BKM Lie superalgebras from
 twisted  CHL dyons", to appear.

\bibitem{9306096}
  T.~Kawai, Y.~Yamada and S.~K.~Yang,
  ``Elliptic Genera And N=2 Superconformal Field Theory,''
  Nucl.\ Phys.\  B {\bf 414}, 191 (1994)
  [arXiv:hep-th/9306096].
  
  \bibitem{dixon}
  L.~J.~Dixon, V.~Kaplunovsky and J.~Louis,
  ``Moduli dependence of string loop corrections to gauge coupling constants,''
  Nucl.\ Phys.\  B {\bf 355}, 649 (1991).



\bibitem{9512046}
  T.~Kawai,
  ``$N=2$ heterotic string 
threshold correction, $K3$ surface and generalized
  Kac-Moody superalgebra,''
  Phys.\ Lett.\  B {\bf 372}, 59 (1996)
  [arXiv:hep-th/9512046].

\bibitem{9607029}
  C.~D.~D.~Neumann,
  ``The elliptic genus of Calabi-Yau 3-folds and 4-folds, product formulae  and
  generalized Kac-Moody algebras,''
  J.\ Geom.\ Phys.\  {\bf 29}, 5 (1999)
  [arXiv:hep-th/9607029].

\end{thebibliography}
\end{document}